\documentclass[twocolumn,twocolappendix]{aastex701}

\usepackage{amsmath}
\usepackage{xcolor}
\graphicspath{{figures/}}

\begin{document}

\title{Mergers Fall Short: Non-merger Channels Required for \\ Galactic Heavy Element Production}

\correspondingauthor{Muhammed Saleem}
\email[show]{muhammed.cholayil@austin.utexas.edu}

\author[orcid=0000-0002-3836-7751, gname=Muhammed,sname=Saleem]{Muhammed Saleem}
\email{muhammed.cholayil@austin.utexas.edu}
\affiliation{Department of Physics, The University of Texas at Austin, 2515 Speedway, Austin, TX 78712, USA}

\author[orcid=0000-0001-5403-3762, gname=Hsin-Yu,sname=Chen]{Hsin-Yu Chen} 
\affiliation{Department of Physics, The University of Texas at Austin, 2515 Speedway, Austin, TX 78712, USA}
\email{hsinyu@austin.utexas.edu}

\author[orcid=0000-0001-6374-6465, gname=Daniel M.,sname=Siegel]{Daniel M. Siegel}
\email{}
\affiliation{Institute of Physics, University of Greifswald, D-17489 Greifswald, Germany}
\affiliation{Department of Physics, University of Guelph, Guelph, Ontario N1G 2W1, Canada}

\author{Philippe Landry}
\email{}
\affiliation{Stripe, Toronto, ON M5J 1T1, Canada}

\author[orcid=0000-0002-3923-1055, gname=Jocelyn S.,sname=Read]{Jocelyn S. Read}
\email{}
\affiliation{Department of Physics, California State University Fullerton, Fullerton, CA, USA}

\author{Kaile Wang}
\email{}
\affiliation{Department of Physics, The University of Texas at Austin, 2515 Speedway, Austin, TX 78712, USA}

\begin{abstract}
Since the discovery of the binary neutron star merger GW170817 and its associated kilonova, neutron star mergers have been established as a key production channel for $r$-process elements in the Universe. However, multiple lines of evidence, including the observations of $r$-process abundances as inferred from stellar spectra of Milky Way disk stars, suggest that additional channels may be needed to fully account for the $r$-process element enrichment in the Milky Way. Neutron star-black hole mergers and fast-merging binary neutron star systems are among the leading alternative candidates. In this paper, we combine gravitational-wave observations from LIGO-Virgo-KAGRA with data from short gamma-ray bursts, Galactic pulsars, and Galactic [Eu/Fe] versus~[Fe/H] abundance observations to assess the contribution of these mergers to $r$-process enrichment in the Galactic disk. 
Our analysis employs a unified, likelihood-based inference framework that consistently propagates uncertainties in merger rates, delay-time distributions, mass- and spin-dependent ejecta yields, and stellar abundance measurements. We find that neither neutron star-black hole mergers nor fast-merging binary neutron star populations can 
dominate the additional prompt enrichment required by the disk-star data unless one invokes extreme assumptions about merger rates or yields. Within the adopted model, these results strongly favor a substantial prompt enrichment component beyond the merger channels considered here.
\end{abstract}


\section{Introduction}\label{sec:intro}

The origins of the heaviest elements in the Universe via rapid neutron capture ($r$-process) nucleosynthesis remains a key question in astrophysics. While it is known that the $r$-process happens in neutron-rich environments \citep{Burbidge1957, Cameron1957}, their dominant astrophysical sites are still debated \citep{Cowan:2019pkx,Siegel:2022upa}. Compact binary mergers, particularly binary neutron star (BNS) mergers, have been recognized as promising sources of $r$-process elements \citep{Lattimer1974, Eichler1989, Freiburghaus1999}. The detection of GW170817 and its associated kilonova provided direct evidence for $r$-process nucleosynthesis in BNS mergers \citep{Abbott2017GW, Abbott2017KN, Kasen2017}. However, the relative role of BNS mergers compared to other potential channels, such as neutron star-black hole (NSBH) mergers \citep{Rosswog2005, Kyutoku2015}, magnetorotational supernovae \citep{Winteler2012, Nishimura2015}, collapsars \cite{Siegel2019}, and giant flares from magnetars~\citep{Cehula:2023hdh,Patel:2025frn} is an open problem. For example, recent chemical evolution studies have raised concerns about whether the event rate evolution of BNS mergers are consistent with the early enrichment seen in metal-poor stars in the Milky Way \citep{Cote2019,Hotokezaka:2018aui,vanDeVoort2015, cescutti2015role,Kobayashi2020,vandeVoort:2021vjz,Kobayashi:2022qlk,Kolborg:2023}. Similar concerns and conclusions are derived from chemical evolution studies of (ultra-faint) dwarf galaxies and globular clusters~\citep{Bonetti:2019fxj,Zevin:2019obe,Skuladottir:2019,Naidu:2021nkt,Simon:2023, Kirby:2023}.

A key observational probe of $r$-process enrichment in the Milky Way is the relationship between $[\mathrm{Eu}/\mathrm{Fe}]$ and $[\mathrm{Fe}/\mathrm{H}]$ (where $[\mathrm{A}/\mathrm{B}]\equiv \log_{10}(N_{\mathrm A}/N_{\mathrm B})-\log_{10}(N_{\mathrm A}/N_{\mathrm B})_{\odot}$; $N$ denotes number abundances and $\odot$ denotes solar), which traces the evolution of europium (Eu, a representative $r$-process element) relative to iron (Fe) as the Galaxy chemically evolves. Observations of metal-poor stars, particularly in the Galactic halo and thick disk, reveal a significant presence of Eu even at low $[\mathrm{Fe}/\mathrm{H}]$ \citep{Sneden2008, Frebel2010}, indicating that $r$-process production was already occurring in the early history of the Galaxy. In order to match the early enrichment, at least some events must occur promptly after star formation---an argument supported by stars like J1521$-$3538, an extremely $r$-process-enhanced star with $[\mathrm{Eu}/\mathrm{Fe}] = +2.2$ and $[\mathrm{Fe}/\mathrm{H}] \sim -3$ \citep{Cain:2020}. The relative abundance of Eu decreases at higher metallicities, implying an increase in late-time Fe production and/or decrease in Eu enrichment. These trends---particularly the negative slope in the [Eu/Fe] versus [Fe/H] relation observed in Milky Way disk stars---place stringent constraints on the rates and yields of $r$-process sources throughout cosmic history. Here, by ``yield'' we mean the per-event $r$-process yield, i.e.\ the total mass of $r$-process material that a single event ejects into the Galactic interstellar medium (often referred to as the nucleosynthesis or chemical yield). Using $[\mathrm{Eu}/\mathrm{Fe}]$--$[\mathrm{Fe}/\mathrm{H}]$ evolution as a reference, one can test whether the proposed astrophysical sites can reproduce the chemical evolution history of the Milky Way.

In our recent study~\citep{HS24}, we investigated whether BNS mergers alone can account for the observed $r$-process enrichment of disk stars in the Milky Way ([Fe/H]$\gtrsim -1$). Standard BNS (``delayed BNS'') formation scenarios predict a delay relative to the star formation rate (SFR)~\citep{Peters:1964,Beniamini:2019iop}. Using a one-zone chemical evolution model \citep{Siegel:2019-Nature} informed by cosmic SFR models~\citep{Madau:2016jbv}, short GRB (sGRB)-inferred delay-time distributions~\citep{Zevin:2022dbo}, GW-inferred BNS rates and mass distributions \citep{Abbott:GWTC3,GWTC3-pop}, neutron star EOS constraints~\citep{Legred:2021hdx}, and semi-analytical ejecta fits \citep{Kruger:2020gig}, we found that BNS mergers following the sGRB-inferred delays cannot reproduce the observed disk-star [Eu/Fe] trend. An additional enrichment channel that tracks the SFR is therefore required, consistent with similar conclusions in the literature~\citep{Cote2019,Kobayashi2020,Kobayashi:2022qlk}. We note that some studies on the metal-poor halo stars and dwarf galaxies have emphasized the importance of delayed r-process sources in ultra-low-metallicity environments (e.g., \cite{Duggan2018NeutronSM} and \cite{Tarumi:2021xvw}). These works probe the earliest phases of chemical enrichment, whereas our analysis focuses on the [Eu/Fe] evolution of Milky Way disk stars at higher metallicities. Relative to \cite{HS24} and related chemical-evolution studies, the aim of the present work is to test quantitatively whether two prominent merger-based  channels---fast-merging BNS systems and NSBH mergers---can satisfy that requirement once current GW-informed merger-rate constraints, ejecta-yield uncertainties, and delay-time information are propagated within one common likelihood framework.

Given the strong evidence for a prompt enrichment channel, we investigate whether two commonly proposed \emph{merger-based} candidates---fast-merging BNS systems and NSBH mergers---can plausibly supply the needed early $r$-process production. Fast-merging BNS may originate from dynamical interactions in dense stellar environments such as globular clusters or galactic nuclei, {although such environments are not expected to produce BNS mergers efficiently enough to dominate over isolated binary evolution}~\citep{Ye:2020gc_bns_rate,Fragione:2020ns_clusters,Belczynski:2018gw170817_origin}, and {are often hypothesized to eject more $r$-process material~\citep{Papenfort:2018:PhysRevD.98.104028,Foucart:2024kci}.} NSBH mergers have similarly been proposed as efficient ejectors under certain mass and spin configurations \citep{Foucart:2018,Mapelli:2018}. Both channels have been suggested as possible solutions to the early-enrichment problem, yet it remains unclear whether either can in fact produce the rates and yields required by the observed [Eu/Fe]--[Fe/H] trends. This is the primary question we resolve here.

Using a likelihood-based inference framework proposed in \cite{HS24}, we assess whether NSBH and fast-merging BNS channels can contribute sufficiently to match the observed evolution of [Eu/Fe] versus [Fe/H], and whether the required conditions are astrophysically plausible given current observational constraints. While the point-estimate studies have shown that standard BNS mergers are unlikely to dominate early Galactic r-process enrichment, such estimates rely on representative values for merger rates, delay times, and ejecta yields. In reality, the r-process yield of compact-binary mergers depends sensitively on the component masses, spins, and equation of state, and the underlying merger population is only weakly constrained by current observations. A likelihood-based inference framework is therefore required to propagate these uncertainties consistently and to determine whether any plausible region of this multi-dimensional parameter space allows merger-based channels to dominate early enrichment. Moreover, point estimates cannot quantify how extreme the required conditions are---for example, whether an additional prompt channel must contribute a modest fraction or more than 95\% of the total r-process budget. By marginalizing over population, yield, and observational uncertainties, our approach provides a quantitative assessment of whether fast-merging BNS or NSBH systems can plausibly serve as the missing prompt enrichment source.

The remainder of this paper is organized as follows. In Methods, we outline the chemical evolution model and describe our inference framework. Appendix~\ref{app:framework} provides a self-contained description of the chemical-evolution equations, rate normalizations, likelihood, and scenario-specific parameterizations used in the analysis. In Results, we discuss results for the three scenarios: fast-merging BNS assumed to be GW-detectable, fast-merging BNS assumed to be GW-undetectable, and NSBH as a secondary channel. Finally, we summarize our findings and discuss their implications for the origin of $r$-process elements in the Milky Way.

\section{Methods}

\subsection{Chemical Evolution Framework with Multiple $r$-process Channels}

Our chemical evolution model, incorporating multiple candidate $r$-process enrichment channels, builds directly upon the one-zone prescription employed in \cite{HS24}, which in turn follows the approach of \cite{Siegel:2019-Nature}. In this framework, the Milky Way is treated as a single, well-mixed reservoir of interstellar medium (ISM). The compact-binary rates used in the inference are volumetric rates, but the chemical-evolution calculation requires Milky-Way-equivalent event rates. We convert between the two using a fiducial number density of Milky-Way-like galaxies, $\rho_{\rm MW}=0.01\,{\rm Mpc}^{-3}$, as described explicitly in Appendix~\ref{app:gce}. This conversion is a normalization choice that allows us to compare local merger-rate densities with the enrichment history of a representative Milky-Way disk. The resulting Milky-Way-equivalent BNS rates are also broadly consistent, within present uncertainties, with independent Galactic double-neutron-star merger-rate estimates from radio pulsar populations \citep{Pol:2020dns_rate,Grunthal:2021dns_rate}. The model evolves the abundances of heavy elements over cosmic time by integrating the $r$-process yields from astrophysical events convolved with their respective rate over time. If the astrophysical events were delayed relative to the SFR, we adopt a delay-time model, which is a power law distribution with minimum delay time ($t_{\rm min}$) and slope ($\alpha$) as parameters (e.g.\ Eq.\ 2 in \cite{Zevin:2022dbo}). For a given SFR, delay-time model, and yields, the model tracks the enrichment history in both $\alpha$ and iron-peak elements, as well as $r$-process elements such as Eu. The full set of abundance-evolution equations and the prompt and delayed rate prescriptions are given in Appendix~\ref{app:framework}.

Key enrichment channels in the baseline model include core-collapse supernovae (CCSNe) and Type Ia supernovae (SNe Ia), mainly contributing to $\alpha$ and iron enrichment. The $r$-process enrichment is primarily attributed to delayed BNS mergers, but the model is flexible enough to include additional candidate channels such as fast-merging BNS systems and NSBH mergers, as we explore in this work. Each channel is parameterized by a local event rate, per-event $r$-process yield or simply yield, and an evolutionary prescription: it either tracks the SFR directly or is obtained by convolving the SFR with a channel-specific delay-time distribution. For comparison with the observed [Eu/Fe]--[Fe/H] trends, we convert the total $r$-process yield of each event into a Eu yield by assuming a fixed $r$-process abundance pattern and adopting the solar $r$-process Eu mass fraction. Thus, unless explicitly stated otherwise, ``yield'' refers to the total $r$-process mass, and Eu yields are obtained as a fixed fraction of this quantity. In all cases, the underlying SFR is taken to be the cosmic star-formation history of \citet{Madau:2016jbv}.

The evolution of the ISM, in addition to all the contributions from various channels, is also governed by a mass-loss term due to star formation and Galactic outflows, which scales with the SFR and which was calibrated in a previous work by \cite{Hotokezaka:2018aui}. For CCSNe and SNe Ia, we follow the same rates and yields for iron production as in~\cite{Siegel:2019-Nature}. In particular, the fiducial SNe Ia prescription adopts a $\tau^{-1}$ delay-time distribution with a minimum delay of $0.4$~Gyr; see Appendix~\ref{app:elements}. This minimum delay is shorter than, or comparable to, many delayed compact-binary merger delays and therefore makes the early [Eu/Fe] decline a stringent test for delayed $r$-process sources. We do not vary the SNe Ia delay-time distribution in the present work, so our quantitative posteriors should be interpreted as conditional on this fixed iron-enrichment model. Thus, the only free parameters in our model are for the $r$-process sources. Given an arbitrary choice of parameters describing each $r$-process channel, the model produces predictions for the time evolution of abundance ratios such as [Fe/H] and [Eu/Fe]. It is this [Eu/Fe] versus~[Fe/H] abundance evolution track that we compare with the observed stellar abundance distributions to constrain possible $r$-process enrichment channels.

\subsection{Likelihood Model and Bayesian Inference Framework}

The stellar abundance data of the disk stars we use are from ~\cite{Battistini}. To quantify the agreement between model predictions and stellar abundance data, we adopt the likelihood formalism introduced in~\cite{HS24}. Each observed disk star is represented as a 2D Gaussian in ([Fe/H], [Eu/Fe]) space, centered on the measured values with widths set by observational uncertainties reported in ~\cite{Battistini}. We use only disk stars, as the one-zone chemical evolution model assumes a well-mixed ISM---an assumption more appropriate for the Galactic disk than for the halo. Halo stars, which formed earlier in more chemically inhomogeneous environments~\citep{Deason:Gaia:2024,Mackereth:2019,Khoperskov:2023,Silva:2024}, are therefore excluded. 
Our use of a one-zone model should therefore be interpreted as a population-level test of the prompt-versus-delayed enrichment requirement in the chemically well-mixed Galactic disk, rather than as an attempt to model local star-to-star abundance scatter or radial abundance gradients. More sophisticated multi-zone, stochastic, or multi-phase ISM treatments can change the detailed mapping between source properties and abundance tracks, especially at the low-metallicity end of the disk sample where the dispersion is non-negligible. For example, two-phase ISM models can modify [Eu/Fe] and [Eu/Mg] evolution by allowing delayed neutron-star-merger ejecta and core-collapse-supernova ejecta to couple differently to the cold star-forming gas \citep{Schonrich:2019two_phase}. Such models may also shift the exact rate and yield values inferred here. However, within the present framework the required fast-merging rates or residual prompt fractions are order-of-magnitude effects, so removing the merger shortfall would require a correspondingly large change in the population-level [Eu/Fe] evolution. Appendix~\ref{app:likelihood} describes how the likelihood should be interpreted under this limitation.

The chemical evolution model described in previous subsection generates a time-ordered track in the same abundance space, and the likelihood for each star is computed by integrating its Gaussian distribution over the model track. The total likelihood is the product over all stars, evaluated numerically. We use this likelihood for Bayesian inference on the parameters governing each $r$-process channel, including the local event rate, yield, and, where applicable, the minimum delay time ($t_{\mathrm{min}}$) and delay-time distribution power-law index ($\alpha$).

For the local merger rates and r-process yields of BNS and NSBH systems, we begin with population inference results of component masses and local merger rates as obtained from the analysis of GWTC-4.0 \citep{GWTC4:population}. These population samples are propagated through the framework of \cite{Chen:2021fro} to obtain joint constraints for the local merger rates and the ejecta masses. In this procedure, neutron-star radii and compactnesses are derived by combining the component masses with independent EOS constraints from joint GW and pulsar observations \citep{Legred:2021hdx}. The resulting compactnesses are then used as inputs to NR-calibrated semi-analytical fitting formulae \citep{Kruger:2020gig} to compute the dynamical ejecta and remnant disk masses with uncertainties as specified in Eq.\ (1) of \cite{HS24}, which are subsequently combined to estimate the total ejecta mass. The resulting joint posterior distributions for the local merger rate and ejecta mass are considered as informative priors in our inference framework. GWTC-5.0 reports no new neutron-star binaries of sufficient significance for the population analysis and yields a lower upper bound on the BNS merger-rate density than GWTC-4.0 \citep{GWTC5:population}. We therefore expect that replacing the GWTC-4.0 rate prior with a GWTC-5.0 prior would strengthen the merger-rate tension reported below. A full rerun with GWTC-5.0 population samples is deferred to future work; it would be required to update the numerical credible intervals quoted in this paper.

Priors on the delay-time distribution parameters are obtained from sGRB constraints~\citep{Zevin:2022dbo}. We use the \texttt{Dynesty} sampler via the \texttt{Bilby} inference framework to sample from the likelihood surface~\citep{Speagle2020dynesty,Ashton:2018jfp}. 
The inference therefore propagates uncertainties in compact-binary rates, delay times, and ejecta yields, while the CCSN and SNe Ia prescriptions are held fixed. Varying the CCSN/SNe Ia rates, Fe yields, or SNe Ia delay-time distribution can shift the detailed abundance track, so our numerical posteriors should be interpreted as conditional on the adopted Fe-production model. Appendix~\ref{app:fe-systematics} presents an explicit sensitivity test showing that the inferred prompt-channel fraction remains close to unity across conservative variations of these fixed Fe-production inputs.

\section{Results}

\subsection{Current GW searches can detect fast-merging BNSs}

We first examine whether a population of fast-merging BNSs can account for the missing $r$-process enrichment {under the assumption that current GW searches are sensitive to both the delayed and fast-merging BNS populations. In this scenario, the GW-inferred BNS merger rate constrains the sum of the two components. We treat the fast-merging channel as an optimistic SFR-tracking limiting case; Appendix~\ref{app:channels} gives the exact rate prescription. A physical fast-merging BNS still requires the stellar-evolution time needed to form two neutron stars and any subsequent inspiral time, so this prescription should not be read as a literal zero-delay model. Rather, it maximizes the early enrichment available from a compact-binary population.}

In this subsection, we discuss the scenario that fast-merging systems are detectable with the same efficiency as delayed BNS mergers. We will discuss the scenario that the fast-merging population is missed from current observations in the next subsection. While these are two extreme scenarios, the reality is likely to be in between the results we find from these two scenarios.

If the fast-merging population is detectable, current GW-inferred BNS merger rates should encompass both delayed and fast-merging systems. The total $r$-process contribution from BNS mergers is then a mixture of both components. We parameterize the fraction $f_{\mathrm{fast}}$ of total BNS mergers that are fast-merging as
\begin{align*}
    \mathrm{R}_\mathrm{fast} &= f_{\mathrm{fast}} \cdot \mathrm{R}_\mathrm{BNS}, \, {\rm and} \\
    \mathrm{R}_\mathrm{delayed} &= (1 - f_{\mathrm{fast}}) \cdot \mathrm{R}_\mathrm{BNS},
\end{align*}
where $\mathrm{R}_\mathrm{BNS}$ is the total BNS rate, $\mathrm{R}_\mathrm{fast}$ is the fast-merging BNS rate, and $\mathrm{R}_\mathrm{delayed}$ is the delayed BNS rate. We assume delayed BNS mergers follow an sGRB-informed delay-time distribution~\citep{Zevin:2022dbo}, while fast-merging BNSs are assumed to be SFR-tracking,  corresponding to the limiting prompt prescription of Eq.~(\ref{eq:app_prompt_rate}). Because of the differing evolution of rates for the two channels, the above rates and $f_{\mathrm{fast}}$ are functions of redshift. We then infer $f_{\mathrm{fast}}$ under this dual-BNS population scenario.

We assume that all BNS populations share the same neutron star EOS and follow the same prescriptions to calculate the $r$-process yields given the component masses~\citep{Kruger:2020gig,HS24}.

\begin{figure}[t]
    \centering
    \includegraphics[width=0.45\textwidth]{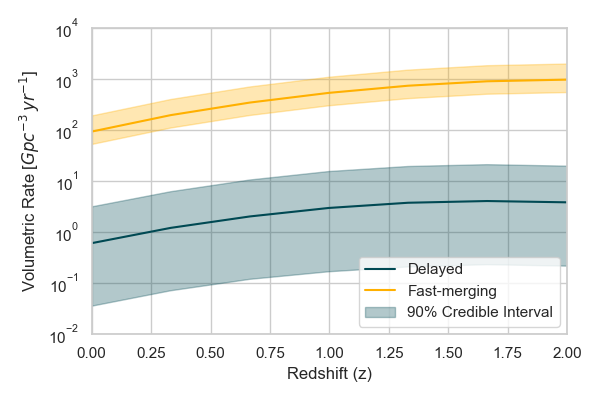}
    \caption{Comoving BNS merger-rate density as a function of redshift for the delayed and fast-merging BNS channels, required if they were solely to explain the [Eu/Fe] versus [Fe/H] abundance trends observed from Milky Way's disk stars. Solid curves are posterior medians; shaded regions show central 90\% credible intervals. Here, we assume the fast-merging systems are detectable by current gravitational-wave searches and have no delay relative to SFR. The figure shows that the fast-merging BNSs have to dominate the BNS rates across the Milky Way history in order to explain the observed abundance trends.}
    \label{fig:ecc_case1_summary}
\end{figure}

Our inference shows that reproducing the observed [Eu/Fe] trend in this scenario requires an exceptionally high fast-merging fraction. The posterior for $f_{\mathrm{fast}}$ in the local Universe ($z=0$) peaks near unity, with more than $98\%$ of BNS mergers inferred to be fast-merging (90\% confidence lower limit). This fraction goes further up at higher redshifts, as a SFR-tracking channel is expected to contribute more frequently in the early Universe compared to delayed sources. In Figure~\ref{fig:ecc_case1_summary}, we show the corresponding volumetric rate as a function of redshift. In particular, the local ($z=0$) volumetric rate of the fast-merging BNS is ${93}_{-41}^{+95}$ Gpc$^{-3}\, \rm{yr}^{-1}$ while the delayed BNS rate remains at a lower level of ${1}_{-1}^{+2}$ Gpc$^{-3}\, \rm{yr}^{-1}$, where the error bars correspond to 90\% credible levels (The reader may refer to the Appendix and Fig.~\ref{fig:fig1_corner_plus_track} for the posteriors of all inferred parameters.). Since this analysis assumes that GW searches are sensitive to both channels, the sum of the two channels is limited by the GW-inferred local BNS rate ($z=0$). It is not the case for our analysis in the next subsection.

These findings suggest that, even with zero delay time, the fast-merging BNSs have to show a near-total dominance in the entire BNS population in order to account for the $r$-process abundance evolution in the Milky Way---specifically the negative slope in [Eu/Fe] versus [Fe/H] for the disk stars. This scenario poses a strong tension with the current understanding of fast-merging BNS populations~\citep{Beniamini:2019iop, Belczynski:2018ptv,Giacobbo:2018etu}, and hence challenges the idea that they are the primary explanation for Galactic $r$-process enrichment.

\subsection{Current GW searches miss fast-merging BNSs}

In this scenario, we consider {a more permissive stress test in which} fast-merging BNSs constitute a distinct subpopulation that is \textit{not} captured by {the GW rate constraint used for delayed BNS mergers}. This {should not be interpreted as a favored astrophysical model or as a claim that current searches are known to miss such a population. Current compact-binary searches include both modeled and weakly modeled approaches, and no observationally established BNS population is known to be hidden at the rates inferred below. Possible reductions in search efficiency could arise from nonstandard signal morphologies, with residual eccentricity in band being one example. Other possible departures include burst-like signals from dynamical capture or near head-on encounters. However, dense stellar environments that can produce eccentric compact binaries are not generally expected to produce BNS mergers efficiently enough to dominate Milky-Way disk enrichment \citep{Ye:2020gc_bns_rate,Fragione:2020ns_clusters,Belczynski:2018gw170817_origin}. Our purpose is therefore to quantify how large such a hidden prompt population would have to be if it existed.}

To model this population, we assume that delayed BNS mergers follow the standard GW-inferred rate and sGRB-informed delay-time distribution~\citep{Zevin:2022dbo}, as in the previous subsection. We then introduce an additional $r$-process channel composed of GW-undetectable fast-merging BNS systems. This additional population is assumed to track the SFR.

The contribution from this channel is parameterized using a multiplicative enhancement factor $X_{\mathrm{fast}}$, such that the total rate from undetectable fast-merging BNS mergers is:
\[
R_{\mathrm{fast}} = X_{\mathrm{fast}} \cdot R_{\mathrm{BNS}},
\]
where $R_{\mathrm{BNS}}$ is the rate of \textit{delayed} BNS mergers. The local ($z=0$) value of $R_{\mathrm{BNS}}$ is inferred from GW observations. Again, the rates and $X_{\mathrm{fast}}$ are functions of redshift.

\begin{figure}[t]
    \centering
    \includegraphics[width=0.45\textwidth]{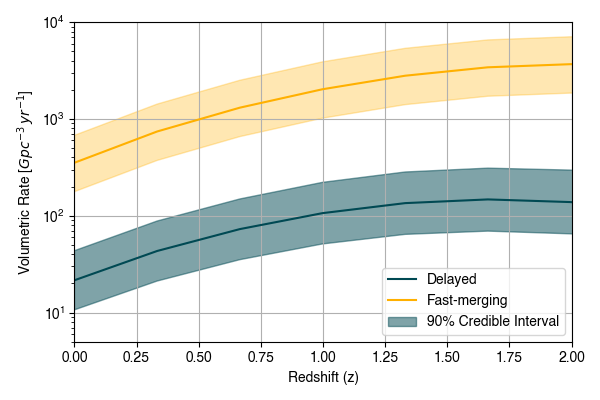}
    \caption{Same as Fig.~\ref{fig:ecc_case1_summary} but under the assumption that the gravitational-wave searches are not sensitive to fast-merging BNS events. The figure shows that a gravitational-wave undetectable fast-merging BNS population can explain the observed $r$-process abundance trends only if their merger rate significantly exceeds that of the delayed population across the Milky Way history.}
    \label{fig:ecc_case2_summary}
\end{figure}

Our analysis finds that, in order to explain the observed [Eu/Fe] versus [Fe/H] trend (negative slope for disk stars, as mentioned before) with only delayed and undetected fast-merging BNSs, the multiplicative factor $X_{\mathrm{fast}}$ is estimated to be ${17.0}_{-6.0}^{+3.0}$ in the local Universe (90\% confidence lower limit), while it further rises up to ${19.0}_{-6.0}^{+6.0}$ at $z = 1$ and ${27.0}_{-10.0}^{+11.0}$ at $z = 2$. Equivalently, the fast-merging BNS rate would be $R_{\mathrm{fast}} = {353.0}_{-176.0}^{+324.0}$ Gpc$^{-3}\, \rm{yr}^{-1}$ in the local Universe (Figure~\ref{fig:ecc_case2_summary}, 90\% confidence lower limit), and $R_{\mathrm{fast}}$ goes up to ${2032.0}_{-1011.0}^{+1867.0}$ Gpc$^{-3}\, \rm{yr}^{-1}$ at $z=1$ and ${3681.0}_{-1832.0}^{+3381.0}$ Gpc$^{-3}\, \rm{yr}^{-1}$ at $z=2$. This implies that the Milky Way would require roughly 17 times (median value) as many fast-merging BNSs as delayed ones at $z=0$; this outcome again challenges the current understanding of the delayed and fast-merging BNS rates~\citep{Beniamini:2019iop, Belczynski:2018ptv,Giacobbo:2018etu} (see the Appendix and Fig.~\ref{fig:fig2_corner_plus_track} for the posteriors of all inferred parameters.).

From the results in the previous and this subsection, we conclude that fast-merging BNSs, even with SFR-tracking rates, are unable to resolve the $r$-process enrichment discrepancy without invoking an extremely high occurrence rate relative to the delayed systems. In both fast-merging scenarios, the posterior support accumulates near the edge of the allowed prompt contribution, indicating that within the adopted model class the data favor even more extreme prompt enrichment than is easily accommodated by plausible compact-binary populations.

The posterior distributions for $f_{\mathrm{fast}}$ and $X_{\mathrm{fast}}$, and the corresponding fast-merging BNS rates derived above quantify the event rates that would be required for fast-merging systems to reproduce the observed Galactic [Eu/Fe] trends within our chemical evolution model. They do not, by themselves, represent the probability that such fast-merging populations exist in nature. A fully self-contained Bayesian hypothesis test of fast-merging-dominated enrichment would require well-calibrated external priors on the intrinsic fast-merging BNS rate, for example from population synthesis, short-GRB statistics, or detailed modeling of dense stellar environments. At present, we lack any widely accepted quantitative constraints on the fast-merging subpopulation in a form that can be robustly incorporated into a second-stage Bayesian model comparison. Our assessment therefore rests on the tension between the exceptionally high fast-merging rates inferred here and current theoretical and observational expectations for compact-binary populations, rather than on a formal statistical exclusion based on externally imposed rate priors.

\subsection{NSBH Mergers as a Secondary Channel}

In this subsection, we assess whether the combined contributions of BNS and NSBH mergers can explain the observed Eu enrichment in the Milky Way. To that end, we construct a joint four-dimensional distribution over the local merger rates and yields of BNS and NSBH systems, following the methodology of~\cite{Chen:2021fro} and using the fitting formulae from~\cite{Kruger:2020gig} for the yield with uncertainties. We use this distribution as priors for our chemical evolution modeling.

As mentioned earlier, the BNS delay-time distribution parameters ($\alpha$, $t_{\min}$) are sampled from sGRB-informed constraints~\citep{Zevin:2022dbo}. The delay time for NSBH mergers is uncertain. It could follow the sGRB delay time, or have a completely different delay-time distribution. If NSBH mergers promptly merge with no delay relative to the SFR, it would contribute most to rectify the difference between the delayed BNS-only evolution track and the [Eu/Fe] versus~[Fe/H] observations, as motivated by \cite{HS24}. We start with this assumption.

We find that even under optimistic delay-time distribution assumptions to maximize the contribution from NSBH, the combined BNS and NSBH channels fail to reproduce the observed [Eu/Fe] trends.

To allow for residual enrichment not captured by BNS and NSBH channels, we then add a third, phenomenological channel whose rate also tracks the SFR. Its contribution is parameterized as a fraction of the total $r$-process enrichment, as its rate and yield are unknown. This component captures any dominant but currently unidentified source of early-time Eu production. It is not intended as a specific progenitor; rather, it functions as a residual prompt component that quantifies the enrichment not supplied by the merger channels considered here.

\begin{figure}[t]
    \centering
    \includegraphics[width=0.45\textwidth]{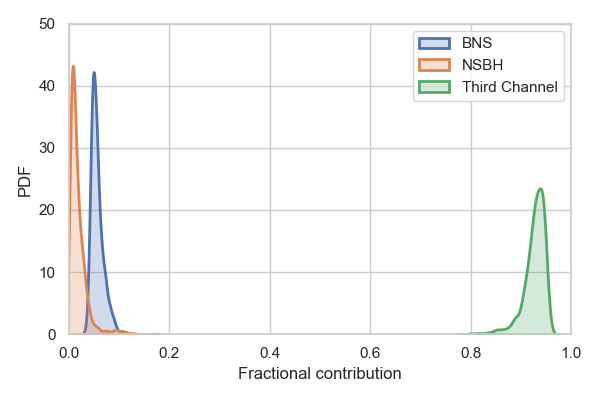}
    \caption{Posterior distributions for the BNS (blue), NSBH (orange), and third-channel (green) $r$-process contributions. We show the fractional contribution from each channel integrated over cosmic history, inferred simultaneously to best fit the observed abundance data from Milky Way disk stars.}
    \label{fig:nsbh_case1_summary}
\end{figure}

Figure~\ref{fig:nsbh_case1_summary} shows the inferred (rate $\times$ yield) fraction integrated over cosmic history until present day, from the BNS (blue), NSBH (orange), and third (green) channels. The BNS contribution is consistent with the findings of \cite{HS24}, while the NSBH contribution remains marginal. In contrast, the third, phenomenological SFR-tracking channel dominates the posterior. Quantitatively, at 90\% confidence, %
within this model space, more than 90\% of the required integrated prompt enrichment is absorbed by the residual third channel, while the combined contribution from BNS and NSBH remains below 10\% (The reader may refer to the Appendix and Fig..~\ref{fig:nsbh_corner_plus_track} for the posteriors of all inferred parameters.). While this does not indicate the existence of a single alternative astrophysical source that contributes exactly this fraction, it can be interpreted as a measure of the merger shortfall within the adopted framework. These results underscore that BNS and NSBH mergers alone cannot account for the observed [Eu/Fe] versus [Fe/H] trend, and that a substantial additional source of $r$-process elements is required.

\section{Discussion and Conclusion}

We have investigated whether fast-merging BNSs or NSBH mergers can resolve the $r$-process enrichment discrepancy identified previously~\citep{Cote2019,Cowan:2019pkx,Kobayashi:2022qlk,vanDeVoort2015,Siegel:2019-Nature,HS24}, where BNS mergers fail to reproduce the observed Eu trends---the negative slope in [Eu/Fe] versus [Fe/H]---in the Milky Way disk. Across all three scenarios explored---GW-detectable fast-merging BNSs, GW-undetectable fast-merging BNSs, and NSBH mergers---we find that, under current theoretical and observational expectations for compact-binary populations, these channels are unlikely to serve as the dominant source of $r$-process enrichment. This conclusion reflects the tension between the event rates required by our chemical-evolution inference and existing expectations for fast-merging and NSBH populations. The central result of this paper is therefore not the identification of a specific alternative source, but a quantitative demonstration that merger-based prompt-enrichment scenarios require extreme rates or yields once current compact-binary constraints are imposed within a common likelihood framework.

Although the sGRB host observations used to infer the delay-time distribution~\citep{Zevin:2022dbo} may suffer from selection biases that favor nearby events, potentially missing fast-merging systems occurring at higher redshifts, our analysis shows that the fast-merging population must be significantly larger than the delayed population---even in the local Universe. {The GW-undetected fast-merger scenario should therefore be read as an intentionally permissive stress test. The required rate enhancement makes this scenario difficult to reconcile with both GW detectability and expected BNS formation rates:}

\begin{enumerate}
\item {First, eccentric or otherwise nonstandard signals can in principle reduce the efficiency of searches based on quasi-circular waveform templates. However, if a hidden population had a local rate $\sim17$ times larger than the delayed BNS rate, current modeled and weakly modeled GW searches~\citep{GWTC4:methods,GWTC5:population} should already constrain at least the nearby, high-signal-to-noise subset. This search-efficiency question can be tested directly with dedicated injection campaigns using eccentric or otherwise nonstandard BNS waveforms.}

\item {Second, mergers with high eccentricity could in principle be produced in dense stellar environments. However, such environments are not expected to produce BNS mergers efficiently enough to dominate over isolated binary evolution: globular-cluster calculations find local NS--NS merger rates of order $0.02~{\rm Gpc}^{-3},{\rm yr}^{-1}$, while young/open-cluster models give rates several orders of magnitude below current empirical GW rates \citep{Ye:2020gc_bns_rate,Fragione:2020ns_clusters}. Hierarchical triples provide another possible pathway, since perturbations from a tertiary companion can excite the eccentricity of the inner binary and enhance gravitational-wave-driven merger \citep{Toonen:2016}. However, the rate of BNS mergers produced through this channel, and the fraction retaining sufficiently high eccentricity to evade standard GW searches, remain uncertain.}

\end{enumerate}

We emphasize that, although the fast-merging BNS and NSBH populations may not precisely follow the assumed star formation rate, may have different yields from those estimated with semi-analytical fit, may have mass distributions that differ from those inferred from GW observations, and/or may not adhere to the assumed $r$-process abundance pattern---all of which could affect their event rates or $r$-process yields---our analysis still indicates that an exceptionally large contribution from these channels is required across cosmic history (see the required fractions and volumetric rates in the \textit{Results} section). This poses a substantial challenge to constructing a consistent theoretical framework. For instance, to reduce the event rates of fast-merging BNSs or NSBHs to a level comparable to that of delayed BNSs, their yield would need to be increased by more than an order of magnitude. This requires a scenario that boosts their yields by an order of magnitude \textit{without increasing the delayed BNS yield}. Similarly, even if the early-Universe evolution of these populations differs from our assumptions, they would still need to occur much more frequently than delayed BNSs to reproduce the observed trends in stellar metallicity. To explore a very different SFR, we repeated our analysis assuming a \textit{constant SFR} and found that our conclusions remain unchanged (see Appendix-\ref{app:sfr} and Fig.~\ref{fig:sfr-effect} for more details).
We also stress that the present calculation does not include metallicity-dependent source formation, metallicity-dependent yields, multi-phase gas evolution, or radial mixing. These effects, including the two-phase ISM effects emphasized by \citet{Schonrich:2019two_phase}, should be included in future chemical-evolution studies, but within the present disk-star, population-level likelihood they would need to substantially change the mean [Eu/Fe] evolution to remove the merger-rate and prompt-fraction tensions found here.

While we have assumed identical r-process yields for delayed and fast-merging BNSs, this assumption is not required. In a more general scenario, the yields from these two channels may differ. In particular, systems classified as fast-merging may preferentially form on highly eccentric orbits, perhaps leading to more violent encounters than quasi-circular binaries and potentially producing a larger mass (25-30\% more) of neutron-rich ejecta, as suggested by previous studies~\citep{Papenfort:2018:PhysRevD.98.104028,Foucart:2024kci}. To explore this possibility, we repeat our fast-merging BNS analyses assuming a yield that is twice that of delayed BNSs (100\% more). Even under this optimistic assumption, our results indicate that fast-merging BNSs must still dominate the total merger rate: they must contribute more than 94\% of the total rate in the scenario where GW searches are sensitive to fast mergers, and require a local-universe rate enhancement of $X_{\rm fast} = 15.0^{+4.0}_{-7.0}$ when GW searches are not sensitive to fast mergers.

We note that this conclusion is also supported by independent astrophysical considerations regarding the environments in which fast-merging BNS systems are expected to form. Such systems are commonly associated with dense stellar environments, including globular clusters and nuclear star clusters. In the Milky Way, globular clusters are predominantly old, metal-poor, and reside in the halo, with star formation ceasing well before the formation of the thin disk considered here, making efficient enrichment of disk stars unlikely. Nuclear star clusters, while more metal-rich, are spatially concentrated near the Galactic center and are similarly disfavored as major contributors to the r-process content of the solar neighborhood. Dynamical calculations of GW170817-like systems also find globular-cluster and nuclear-cluster channels to be subdominant to isolated binary evolution \citep{Belczynski:2018gw170817_origin}. These considerations provide complementary physical support for our conclusion that fast-merging BNSs are unlikely to dominate r-process enrichment in the Galactic thin disk.

Future extensions should also include additional abundance ratios, such as [Eu/Mg] or other alpha-element tracers, because non-compact-binary prompt sources may synthesize alpha elements in addition to $r$-process material. Such information could help distinguish a compact-binary prompt channel from alternatives such as rare classes of core-collapse explosions.

Our findings not only reinforce the conclusions of \cite{HS24}, but also pose substantial challenges to two leading candidates---fast-merging BNS and NSBH systems---as dominant $r$-process enrichment channels. These results help narrow down the viable $r$-process sources in the Milky Way and strongly favor an additional prompt enrichment component beyond the merger channels considered here, although its physical origin remains to be established.

\begin{acknowledgments}
The authors would like to thank Joon Young Lee, Floor Broekgaarden, Wen-fai Fong, Francois Foucart, Erika Holmbeck, and Alexander Ji for very helpful discussions. M.S.\ is supported by the Weinberg Institute for Theoretical Physics at the University of Texas at Austin. H.-Y.C. is supported by the National Science Foundation under Grant PHY-2308752 and Department of Energy Grant DE-SC0025296. The authors are grateful for computational resources provided by the LIGO Laboratory and supported by National Science Foundation Grants PHY-0757058 and PHY-0823459. This material is based upon work supported by NSF's LIGO Laboratory which is a major facility fully funded by the National Science Foundation.
\end{acknowledgments}

\appendix

\section{Chemical-evolution and Bayesian inference framework}
\label{app:framework}

In this section, we provide the details of the chemical-evolution and
statistical framework used in this work. The framework combines fixed
prescriptions for hydrogen and iron enrichment with parameterized
$r$-process source populations, specified by their event-rate histories,
delay-time distributions, and ejecta yields. For any chosen set of source
parameters, the calculation produces a Galactic abundance track in the
$([\mathrm{Fe}/\mathrm{H}],[\mathrm{Eu}/\mathrm{Fe}])$ plane. This track
is then compared directly with stellar-abundance measurements through a
likelihood defined in the same two-dimensional abundance space.

\subsection{One-zone chemical-evolution model}
\label{app:gce}

We model the Milky Way as a single, well-mixed reservoir of interstellar
gas. The abundance evolution is initialized at redshift $z=10$ and
followed to the present time. At earlier epochs the adopted star-formation
history contributes negligibly to the enrichment modeled here. We adopt a
flat cosmology with $\Omega_{\rm M}=0.3$, $\Omega_{\Lambda}=0.7$, and
$h=0.7$ to convert between redshift and cosmic time. These cosmological
parameters provide the common time coordinate used for the star-formation
history and delayed source populations.

The fiducial star-formation-rate density is taken from
\citet{Madau:2016jbv},
\begin{equation}
    \psi(z)
    =
    10^{9}
    \frac{0.01(1+z)^{2.6}}
    {1+\left[(1+z)/3.2\right]^{6.2}}
    \;
    M_{\odot}\,
    {\rm Gyr}^{-1}\,
    {\rm Mpc}^{-3},
    \label{eq:app_sfr}
\end{equation}
where the factor of $10^{9}$ converts the conventional
$M_{\odot}\,{\rm yr}^{-1}\,{\rm Mpc}^{-3}$ normalization to
$M_{\odot}\,{\rm Gyr}^{-1}\,{\rm Mpc}^{-3}$. We convert volumetric
quantities into Milky-Way-equivalent quantities using the number density
of Milky-Way-like galaxies
\begin{equation}
    \rho_{\rm MW}=0.01\ {\rm Mpc}^{-3}
\end{equation}
\citep{Abadie2010}. The corresponding star-formation rate of a
Milky-Way-equivalent galaxy is
\begin{equation}
    \psi_{\rm MW}(t)
    =
    \frac{\psi(t)}{\rho_{\rm MW}}.
    \label{eq:app_mw_sfr}
\end{equation}

Following the one-zone prescription of
\citet{Siegel:2019-Nature} and \citet{HS24}, the interstellar-medium
mass is related to the star-formation rate through a
Kennicutt--Schmidt-type scaling,
\begin{equation}
    M_{\rm ISM}(t)
    =
    10^{10}\,M_{\odot}
    \left[
        \frac{\psi_{\rm MW}(t)}
        {\psi_{\rm MW}(t_0)}
    \right]^{1/1.4},
    \label{eq:app_ism}
\end{equation}
where $t_0$ is the present cosmic time. The hydrogen mass is
\begin{equation}
    M_{\rm H}(t)
    =
    f_{\rm H}M_{\rm ISM}(t),
    \qquad
    f_{\rm H}=0.75,
    \label{eq:app_hydrogen}
\end{equation}
where $f_{\rm H}$ is the assumed hydrogen mass fraction of the ISM. We keep $f_{\rm H}$ fixed as a bookkeeping approximation for converting the evolving gas reservoir into a hydrogen mass; gas consumption and outflows are instead represented through the evolution of $M_{\rm ISM}(t)$ and through the proportional loss terms below.

Star formation locks gas into stars, while Galactic outflows remove gas
from the system. In a one-zone, well-mixed model, both processes remove
each element in proportion to its instantaneous mass fraction. For a
generic element $Z$, the corresponding loss term is
\begin{equation}
    \left.
    \frac{dM_Z}{dt}
    \right|_{\rm loss}
    =
    -\Lambda(t)M_Z(t),
    \label{eq:app_generic_loss}
\end{equation}
where
\begin{equation}
    \Lambda(t)
    =
    (1+\eta)
    \frac{\psi_{\rm MW}(t)}
    {M_{\rm ISM}(t)},
    \qquad
    \eta=0.25.
    \label{eq:app_loss}
\end{equation}
The first term in $1+\eta$ represents removal by star formation and
$\eta$ is the outflow mass-loading factor, defined such that the outflow
rate is $\eta\psi_{\rm MW}$. This prescription follows the one-zone
treatment used in \citet{Hotokezaka:2018aui},
\citet{Siegel:2019-Nature}, and \citet{HS24}. Because the ISM is assumed
to be instantaneously mixed, Eq.~(\ref{eq:app_generic_loss}) removes all
tracked elements in proportion to their current abundances.

\subsection{Prompt and delayed enrichment channels}
\label{app:channels}

Each enrichment source $i$ is characterized by an event-rate history
$R_i(t)$ and a per-event yield $m_{Z,i}$ of the chemical species or
nucleosynthetic component being tracked. The framework accommodates both
sources whose event rates follow the star-formation history and sources
whose events occur after a distribution of delays relative to the
formation of their progenitor systems.

For a delayed source, we adopt a power-law delay-time distribution,
\begin{equation}
    D_i(\tau)
    =
    C_i\,\tau^{\alpha_i}
    \Theta\left(\tau-\tau_{{\rm min},i}\right),
    \label{eq:app_dtd}
\end{equation}
where $\tau$ is the time between progenitor formation and the enrichment
event, $\alpha_i$ is the power-law index,
$\tau_{{\rm min},i}$ is the minimum delay time, and $\Theta$ is the
Heaviside function. The corresponding volumetric event-rate density is
\begin{equation}
    R_i(t)
    =
    \int_{t_{\rm start}}^{t}
    \psi(t')D_i(t-t')\,dt',
    \label{eq:app_delayed_rate}
\end{equation}
where $t_{\rm start}\equiv t(z=10)$ is the initial time of the
chemical-evolution calculation.

When a channel is parameterized by its present-day local rate,
$R_{i,0}\equiv R_i(t_0)$, the DTD normalization is
\begin{equation}
    C_i
    =
    \frac{R_{i,0}}
    {
        \displaystyle
        \int_{t_{\rm start}}^{t_0}
        \psi(t')
        (t_0-t')^{\alpha_i}
        \Theta\left(
            t_0-t'-\tau_{{\rm min},i}
        \right)
        \,dt'
    }.
    \label{eq:app_dtd_norm}
\end{equation}
This choice guarantees that the convolved rate agrees with the specified
local rate at $t=t_0$.

A source that tracks the star-formation rate is instead assigned
\begin{equation}
    R_i(t)
    =
    R_{i,0}
    \frac{\psi(t)}{\psi(t_0)}.
    \label{eq:app_prompt_rate}
\end{equation}
Throughout this paper, ``prompt,'' ``SFR-tracking,'' and, in the
fast-merging analyses, ``fast-merging'' refer to this adopted limiting
prescription. A physically fast-merging binary could have a short but
nonzero delay; here we use the SFR-tracking limit because it maximizes
the ability of such a population to enrich the Galaxy at early times.
It should therefore be understood as an optimistic limiting case, not
as a claim that every event occurs at exactly zero delay.

The corresponding Milky-Way-equivalent event rate is
\begin{equation}
    R_{i,{\rm MW}}(t)
    =
    \frac{R_i(t)}{\rho_{\rm MW}}.
    \label{eq:app_mw_rate}
\end{equation}

\subsection{Iron and europium enrichment}
\label{app:elements}

The iron history is supplied by core-collapse supernovae (CCSNe) and
Type Ia supernovae (SNe Ia). CCSNe are taken to follow the
star-formation rate, while SNe Ia follow a delayed rate with a
power-law DTD. These prescriptions and the associated iron yields follow
the chemical-evolution model of \citet{Siegel:2019-Nature}; the
$\tau^{-1}$ Type Ia DTD is also consistent with the observationally
inferred form summarized by \citet{Maoz:2017yjc}. Their rates and yields
are held fixed because they are substantially better constrained than
the compact-binary $r$-process rates and yields explored here, and their
remaining uncertainties are not expected to drive the conclusions of
this analysis.

The iron mass evolves according to
\begin{align}
    \frac{dM_{\rm Fe}}{dt}
    ={}&
    m_{\rm Fe}^{\rm CCSN}
    R_{{\rm CCSN},{\rm MW}}(t)
    +
    m_{\rm Fe}^{\rm Ia}
    R_{{\rm Ia},{\rm MW}}(t)
    \nonumber\\
    &-
    \Lambda(t)M_{\rm Fe}(t),
    \label{eq:app_fe_ode}
\end{align}
where
\begin{equation}
    m_{\rm Fe}^{\rm CCSN}=0.074\,M_{\odot},
    \qquad
    m_{\rm Fe}^{\rm Ia}=0.7\,M_{\odot}.
    \label{eq:app_fe_yields}
\end{equation}
These yields are the values adopted in the baseline model of
\citet{Siegel:2019-Nature}. The Type Ia rate uses
$D_{\rm Ia}(\tau)\propto\tau^{-1}$ with a minimum delay time of
$0.4$ Gyr, following the fiducial value adopted in the baseline
prescription of \citet{Siegel:2019-Nature}; empirical SNe Ia DTDs are
commonly described by an approximately $\tau^{-1}$ form
\citep{Maoz:2017yjc}. Together, these fixed inputs
determine the evolution of $M_{\rm Fe}(t)$ and hence the horizontal
coordinate $[\mathrm{Fe}/\mathrm{H}](t)$ of the abundance track.

For an $r$-process source $i$, the Eu mass injected per event is obtained
from its total $r$-process ejecta mass,
\begin{equation}
    m_{{\rm Eu},i}
    =
    f_{{\rm Eu},r}\,
    m_{{\rm ej},i},
    \label{eq:app_eu_yield}
\end{equation}
where
\begin{equation}
    f_{{\rm Eu},r}
    =
    1.04\times10^{-3}
    \label{eq:app_eu_fraction}
\end{equation}
is the Eu mass fraction of the adopted solar $r$-process residual
pattern \citep{Arnould:2007gh}. Thus, unless explicitly stated otherwise,
``yield'' in this paper refers to the total $r$-process ejecta mass,
whereas the Eu yield used in the chemical-evolution equations is the
fixed fraction given in Eq.~(\ref{eq:app_eu_yield}).

The Eu source term from channel $i$ is
\begin{equation}
    \dot{M}_{{\rm Eu},i}(t)
    =
    f_{{\rm Eu},r}\,
    m_{{\rm ej},i}\,
    R_{i,{\rm MW}}(t).
    \label{eq:app_eu_source}
\end{equation}
For an arbitrary collection of $r$-process channels, the Eu mass evolves
as
\begin{align}
    \frac{dM_{\rm Eu}}{dt}
    =
    \sum_i
    f_{{\rm Eu},r}\,
    m_{{\rm ej},i}\,
    R_{i,{\rm MW}}(t)
    -
    \Lambda(t)M_{\rm Eu}(t).
\label{eq:app_eu_ode}
\end{align}

For any specified star-formation history and source model, the coupled
equations above determine $M_{\rm H}(t)$, $M_{\rm Fe}(t)$, and
$M_{\rm Eu}(t)$. The first two are fixed once the star-formation history
and the CCSN/SN Ia prescriptions are chosen, while $M_{\rm Eu}(t)$
depends on the rates, yields, and, where relevant, delay-time parameters
of the adopted $r$-process channels. Repeating the analysis with a
different star-formation history changes all of these histories
consistently.

The model masses are converted to the logarithmic number-abundance
ratios used in the stellar data:
\begin{align}
    [\mathrm{Fe}/\mathrm{H}](t)
    ={}&
    \log_{10}
    \left[
        \frac{M_{\rm Fe}(t)}{M_{\rm H}(t)}
        \frac{A_{\rm H}}{A_{\rm Fe}}
    \right]
    -
    \log_{10}
    \left(
        \frac{N_{\rm Fe}}{N_{\rm H}}
    \right)_{\odot},
    \label{eq:app_feh}
    \\
    [\mathrm{Eu}/\mathrm{Fe}](t)
    ={}&
    \log_{10}
    \left[
        \frac{M_{\rm Eu}(t)}{M_{\rm Fe}(t)}
        \frac{A_{\rm Fe}}{A_{\rm Eu}}
    \right]
    -
    \log_{10}
    \left(
        \frac{N_{\rm Eu}}{N_{\rm Fe}}
    \right)_{\odot}.
    \label{eq:app_eufe}
\end{align}
We use $A_{\rm H}=1.008$, $A_{\rm Fe}=55.845$, and
$A_{\rm Eu}=151.96$. The adopted solar photospheric abundances are
\begin{equation}
    \log_{10}
    \left(
        \frac{N_{\rm Fe}}{N_{\rm H}}
    \right)_{\odot}
    =-4.50,
    \qquad
    \log_{10}
    \left(
        \frac{N_{\rm Eu}}{N_{\rm H}}
    \right)_{\odot}
    =-11.48
    \label{eq:app_solar_abundances}
\end{equation}
\citep{Asplund2009}. Consequently,
\begin{equation}
    \log_{10}
    \left(
        \frac{N_{\rm Eu}}{N_{\rm Fe}}
    \right)_{\odot}
    =-6.98.
    \label{eq:app_solar_eufe}
\end{equation}

For a generic parameter vector
\begin{equation}
    \boldsymbol{\theta}
    =
    \left\{
    R_{i,0},
    m_{{\rm ej},i},
    \alpha_i,
    \tau_{{\rm min},i},
    \ldots
    \right\},
    \label{eq:app_generic_theta}
\end{equation}
with only the subset relevant to the adopted source model included, the
calculation produces the time-ordered abundance track
\begin{equation}
    \boldsymbol{y}(t;\boldsymbol{\theta})
    =
    \left(
        [\mathrm{Fe}/\mathrm{H}](t),
        [\mathrm{Eu}/\mathrm{Fe}](t)
    \right).
    \label{eq:app_track}
\end{equation}

\subsection{Likelihood for the stellar abundances}
\label{app:likelihood}

We compare the predicted abundance track with the Milky Way disk-star
measurements of \citet{Battistini}. We restrict the analysis to disk
stars because the one-zone, well-mixed approximation is more appropriate
for the Galactic disk than for the chemically inhomogeneous halo. The
adopted data set contains $N_{\star}=378$ stars.

For star $j$, let
\begin{equation}
    \boldsymbol{x}_j
    =
    \left(
        [\mathrm{Fe}/\mathrm{H}]_j,
        [\mathrm{Eu}/\mathrm{Fe}]_j
    \right)
\end{equation}
denote the measured abundances. We represent its measurement uncertainty
by the two-dimensional Gaussian
\begin{equation}
    p_j(\boldsymbol{x})
    =
    \mathcal{N}
    \left(
        \boldsymbol{x};
        \boldsymbol{x}_j,
        \boldsymbol{\Sigma}_j
    \right),
    \label{eq:app_star_gaussian}
\end{equation}
with
\begin{equation}
    \boldsymbol{\Sigma}_j
    =
    \begin{pmatrix}
        \sigma_{{\rm FeH},j}^{2} & 0 \\
        0 & \sigma_{{\rm EuFe},j}^{2}
    \end{pmatrix}.
    \label{eq:app_covariance}
\end{equation}
The adopted uncertainties are
$\sigma_{\rm FeH}=0.05$ dex and
$\sigma_{\rm EuFe}=0.094$ dex.

Because the chemical-evolution model predicts a curve rather than a
single point in the abundance plane, the likelihood contribution of
star $j$ is obtained by integrating its Gaussian density along the
model track:
\begin{equation}
    \mathcal{L}_j(\boldsymbol{\theta})
    =
    \int_{\rm track}
    \mathcal{N}
    \left[
        \boldsymbol{y};
        \boldsymbol{x}_j,
        \boldsymbol{\Sigma}_j
    \right]
    d[\mathrm{Fe}/\mathrm{H}].
    \label{eq:app_star_likelihood}
\end{equation}
The integral is evaluated numerically using trapezoidal quadrature over
the ordered track points. This construction does not assign a formation
time to an individual star; it asks whether the measured abundance lies
near any portion of the chemical-evolution trajectory.

Assuming the stellar measurements are independent, the total abundance
likelihood is
\begin{equation}
    \ln\mathcal{L}_{\rm abund}
    \left(
        \mathcal{D}_{\star}
        \mid
        \boldsymbol{\theta}
    \right)
    =
    \sum_{j=1}^{N_{\star}}
    \ln\mathcal{L}_j(\boldsymbol{\theta}).
    \label{eq:app_total_likelihood}
\end{equation}

The one-zone approximation is not intended to reproduce the
star-to-star scatter generated by stochastic or spatially inhomogeneous
enrichment, particularly at low metallicity. The likelihood instead
tests whether the population-level evolution of the disk abundances can
be reproduced by the assumed combination of prompt and delayed sources.

\subsection{Astrophysical constraints and posterior inference}
\label{app:bayesian}

The abundance likelihood is combined with external information about
compact-binary populations and delay times. Denoting the complete set of
external constraints by $\mathcal{D}_{\rm ext}$, the posterior has the
generic form
\begin{equation}
    p
    \left(
        \boldsymbol{\theta}
        \mid
        \mathcal{D}_{\star},
        \mathcal{D}_{\rm ext}
    \right)
    \propto
    \mathcal{L}_{\rm abund}
    \left(
        \mathcal{D}_{\star}
        \mid
        \boldsymbol{\theta}
    \right)
    \pi_{\rm ext}(\boldsymbol{\theta})
    \pi_{\rm model}(\boldsymbol{\theta}),
    \label{eq:app_posterior}
\end{equation}
where $\pi_{\rm ext}$ represents the informative constraints obtained
from GW and sGRB observations, and $\pi_{\rm model}$ contains the
remaining priors specific to the enrichment scenario.

For compact-binary channels, we begin with population-inference samples
for the local merger rates and component masses from GW observations
\citep{GWTC4:population}. Following \citet{Chen:2021fro}, the component
masses are combined with neutron-star equation-of-state constraints
\citep{Legred:2021hdx} to obtain radii and compactnesses. These
quantities are propagated through numerical-relativity-calibrated
fitting formulae \citep{Kruger:2020gig} to estimate the dynamical ejecta
and remnant-disk masses. The result is a joint distribution of local
merger rates and total $r$-process ejecta yields, with non-zero correlations. For the delayed BNS component, the DTD parameters
$(\alpha,\tau_{\min})$ are constrained by the sGRB-informed posterior of
\citet{Zevin:2022dbo}. 

Posterior sampling is performed with \texttt{Dynesty} through the
\texttt{Bilby} inference framework
\citep{Speagle2020dynesty,Ashton:2018jfp}. For each proposed
$\boldsymbol{\theta}$, we construct the channel-rate histories, compute
the Eu injection and abundance track, evaluate
Eq.~(\ref{eq:app_total_likelihood}), and combine the result with the
external and model-specific prior factors in
Eq.~(\ref{eq:app_posterior}). The hydrogen and iron histories are fixed
within a run with a specified star-formation model and are therefore
computed once, whereas the Eu history is recomputed as the
$r$-process parameters vary. When the analysis is repeated with a
different star-formation history, all fixed chemical-evolution
quantities are recomputed consistently.

\subsection{Parameterization of the three analyses}
\label{app:three_cases}

The chemical-evolution and likelihood framework above is common to the
three analyses. They differ only in the parameterization of their
$r$-process source populations.

\subsubsection{Fast-merging BNS systems included in current GW searches}

In the first analysis, the GW-inferred BNS rate is interpreted as the
total rate of delayed and fast-merging BNS systems. Their local rates are
\begin{align}
    R_{{\rm fast},0}
    &=
    f_{\rm fast}R_{{\rm BNS},0},
    \\
    R_{{\rm delayed},0}
    &=
    (1-f_{\rm fast})R_{{\rm BNS},0}.
    \label{eq:app_detectable_split}
\end{align}
The delayed component follows
Eq.~(\ref{eq:app_delayed_rate}), with its DTD constrained by sGRB
observations, while the fast component follows
Eq.~(\ref{eq:app_prompt_rate}). The two components share the same
baseline ejecta-yield prescription unless explicitly stated otherwise.
The analysis samples the total local BNS rate, the common ejecta yield,
the delayed-channel DTD parameters, and $f_{\rm fast}$.

\subsubsection{Fast-merging BNS systems missed by current GW searches}

In the second analysis, the GW-inferred BNS population is identified
with the delayed component alone. An additional fast-merging population
is introduced with
\begin{equation}
    R_{{\rm fast},0}
    =
    X_{\rm fast}R_{{\rm delayed},0}.
    \label{eq:app_undetectable_scaling}
\end{equation}
The delayed population follows the sGRB-informed DTD, while the added
fast population tracks the SFR. The analysis samples the delayed BNS
rate, the common ejecta yield, the delayed-channel DTD parameters, and
$X_{\rm fast}$. Because the fast population is assumed to be absent from
the current GW population inference, the sum of the delayed and fast
rates is not restricted to the GW-inferred BNS rate.

\subsubsection{BNS, NSBH, and residual prompt enrichment}

In the third analysis, delayed BNS mergers and SFR-tracking NSBH mergers
are included simultaneously. Because the chemical enrichment depends on
the product of event rate and ejecta mass, we define
\begin{equation}
    \mathcal{Y}_i
    \equiv
    R_{i,0}m_{{\rm ej},i},
    \qquad
    i\in\{\mathrm{BNS},\mathrm{NSBH}\}.
    \label{eq:app_yield_rate}
\end{equation}
The joint prior on
$(\mathcal{Y}_{\rm BNS},\mathcal{Y}_{\rm NSBH})$ is obtained by
propagating the GW population, equation-of-state, and ejecta-model
uncertainties described above.

To quantify any enrichment not supplied by the two merger channels, we
include an additional phenomenological SFR-tracking component. Its
present-day rate--yield contribution is parameterized as
\begin{equation}
    \mathcal{Y}_3
    =
    \frac{f_3}{1-f_3}
    \left(
        \mathcal{Y}_{\rm BNS}
        +
        \mathcal{Y}_{\rm NSBH}
    \right),
    \label{eq:app_third_channel}
\end{equation}
so that
\begin{equation}
    f_3
    =
    \frac{\mathcal{Y}_3}
    {
        \mathcal{Y}_{\rm BNS}
        +
        \mathcal{Y}_{\rm NSBH}
        +
        \mathcal{Y}_3
    }.
    \label{eq:app_f3}
\end{equation}
The sampled parameters are the BNS and NSBH rate--yield products, the
BNS DTD parameters, and $f_3$. The third component is not identified
with a specific progenitor; it measures the residual prompt enrichment
required after accounting for the BNS and NSBH contributions. The
integrated channel fractions reported in the main text are obtained by
integrating the corresponding mass-injection histories over cosmic time.

\subsection{Interpretation and fixed assumptions}
\label{app:interpretation}

The inference varies the rates, yields, delay-time parameters, and
channel fractions associated with the $r$-process sources while holding
fixed the one-zone mixing prescription, gas outflow model, CCSN and
Type Ia enrichment, solar abundance scale, and conversion between total
$r$-process ejecta and Eu. The fiducial analysis also fixes the cosmic
star-formation history, while Appendix~\ref{app:sfr} explicitly tests
the effect of replacing it with a constant star-formation history.

The resulting posteriors should therefore be interpreted as conditional
on this chemical-evolution framework. The purpose of the analysis is not
to infer every aspect of Galactic chemical evolution simultaneously,
but to determine whether the uncertainties in compact-binary rates,
delays, and ejecta yields are sufficient for the merger populations
considered here to reproduce the observed disk-star abundance trend.

\section{Sensitivity to fixed iron-enrichment prescriptions}
\label{app:fe-systematics}

We varied the CCSN local rate over $7.83\times10^{-5}$--$1.066\times10^{-4}~{\rm yr}^{-1}~{\rm Mpc}^{-3}$, motivated by the Palomar Transient Factory CCSN rate measurement of \citet{Frohmaier:2021ccsn_rate}, and the CCSN Fe yield over $0.051$--$0.065~M_{\odot}$, corresponding to the $0.058\pm0.007~M_{\odot}$ empirical estimate of \citet{Rodriguez:2023ccsn_fe}. For SNe Ia, we varied the time-integrated DTD normalization over $(1.2$--$1.4)\times10^{-3}~M_{\odot}^{-1}$ around the \citet{Maoz:2017yjc} value, the Fe yield over $0.65$--$0.75~M_{\odot}$, and repeated the corner cases for both the fiducial minimum delay $t_{\rm Ia,min}=0.4~{\rm Gyr}$ and a prompt value $t_{\rm Ia,min}=0.04~{\rm Gyr}$ commonly used when integrating empirical SNe Ia DTDs. The four Fe-rate/yield corners are: all CCSN and SNe Ia inputs high; all inputs low; CCSN inputs high with SNe Ia inputs low; and CCSN inputs low with SNe Ia inputs high.

\begin{figure}[t]
    \centering
    \includegraphics[width=\columnwidth]{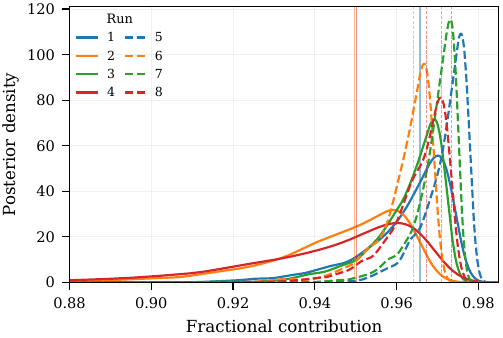}
    \caption{Posterior distributions for the fractional contribution of the unknown prompt $r$-process channel in the Fe-enrichment sensitivity tests. Runs 1--4 use $t_{\rm Ia,min}=0.4~{\rm Gyr}$ and runs 5--8 use $t_{\rm Ia,min}=0.04~{\rm Gyr}$. Within each group, the scenarios are: (1,5) all CCSN and SNe Ia Fe-production inputs set to their high values; (2,6) all set to their low values; (3,7) CCSN inputs high and SNe Ia inputs low; and (4,8) CCSN inputs low and SNe Ia inputs high. Solid and dashed curves therefore compare the same Fe-rate/yield corner under the two SNe Ia minimum-delay choices. Across all eight cases, the posterior remains concentrated at large prompt fractions, with median values $f_{\rm prompt}\simeq0.95$--$0.97$. Thus, varying the fixed Fe-production inputs within these ranges does not remove the need for a dominant prompt enrichment component.}
    \label{fig:fe-systematics}
\end{figure}

The resulting posteriors are shown in Fig.~\ref{fig:fe-systematics}. The median prompt-channel fraction ranges from $0.95$ to $0.97$ across all eight stress tests, and even the 5th percentile remains above $0.90$ in every case. The Fe prescriptions therefore shift the detailed abundance-track calibration but do not change the qualitative conclusion: within this one-zone framework, delayed BNS enrichment cannot supply most of the required Galactic $r$-process budget, and a dominant prompt component is still required.


\section{Additional details for all three cases}\label{sec:Appendix}

\begin{figure*}
    \centering
    \includegraphics[width=0.75\textwidth]{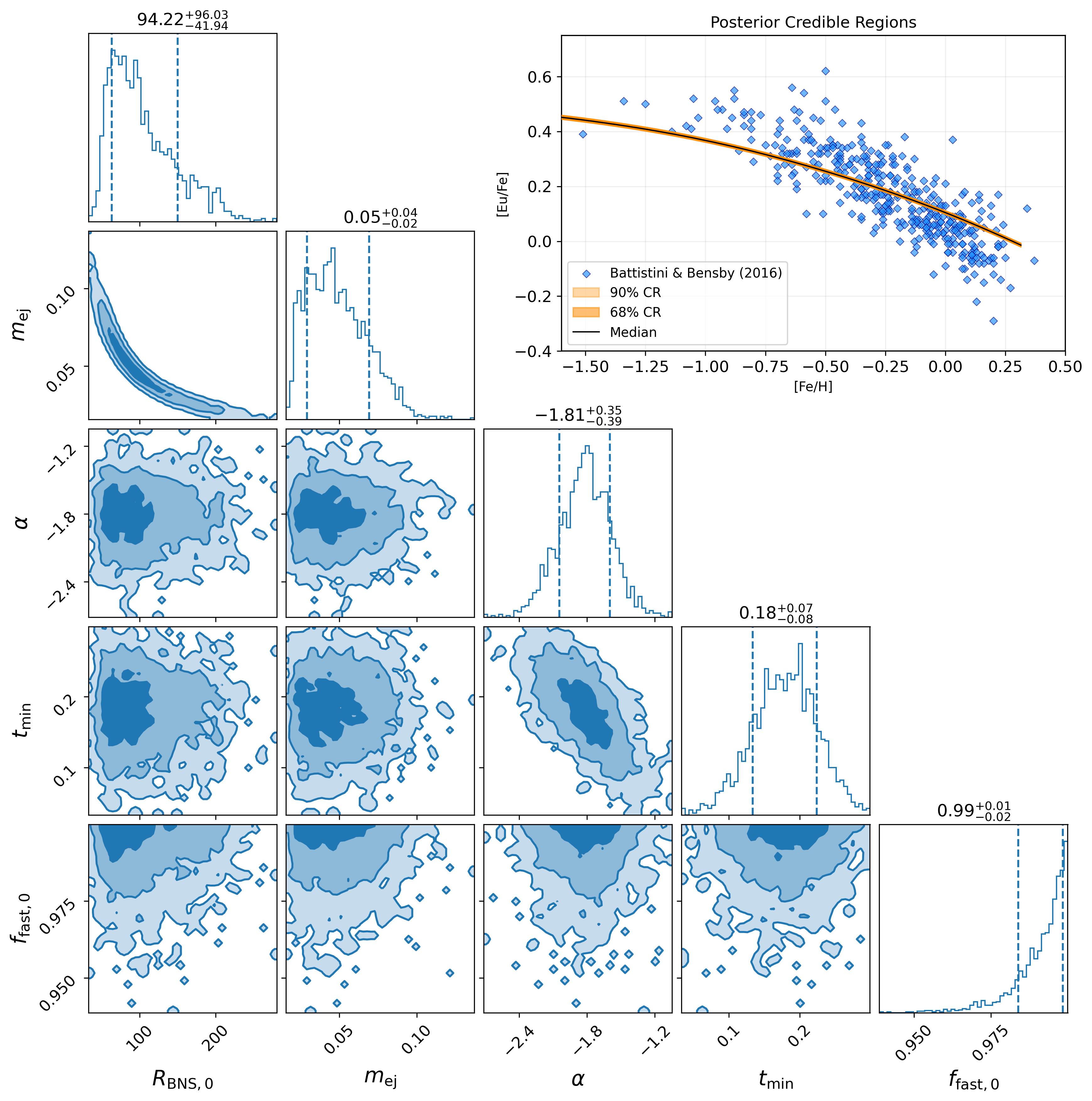}
    \caption{\textbf{Current GW searches can detect fast-merging BNSs.} Diagonal panels show marginalized posteriors (medians and 68\% credible intervals); off-diagonal panels show joint posteriors (filled credible regions). Parameters are the total BNS rate $R_{\rm BNS}$ (${\rm Gpc}^{-3}\, {\rm yr}^{-1}$), ejecta mass $m_{\rm ej}$ (${\rm M}_\odot$), delayed-channel delay-time parameters $\alpha$ and $t_{\rm min}$, and the fast-merging fraction $f_{\rm fast}$. The subscript `0' in the parameter labels indicate that the quantities are evaluated in the local universe ($z=0$). The large panel at upper right displays the posterior-predicted credible regions (orange) in $[\mathrm{Eu}/\mathrm{Fe}]$ versus $[\mathrm{Fe}/\mathrm{H}]$, with Milky Way disk measurements from \cite{Battistini} shown as blue points.}
    \label{fig:fig1_corner_plus_track}
\end{figure*}

\begin{figure*}
    \centering
    \includegraphics[width=0.75\textwidth]{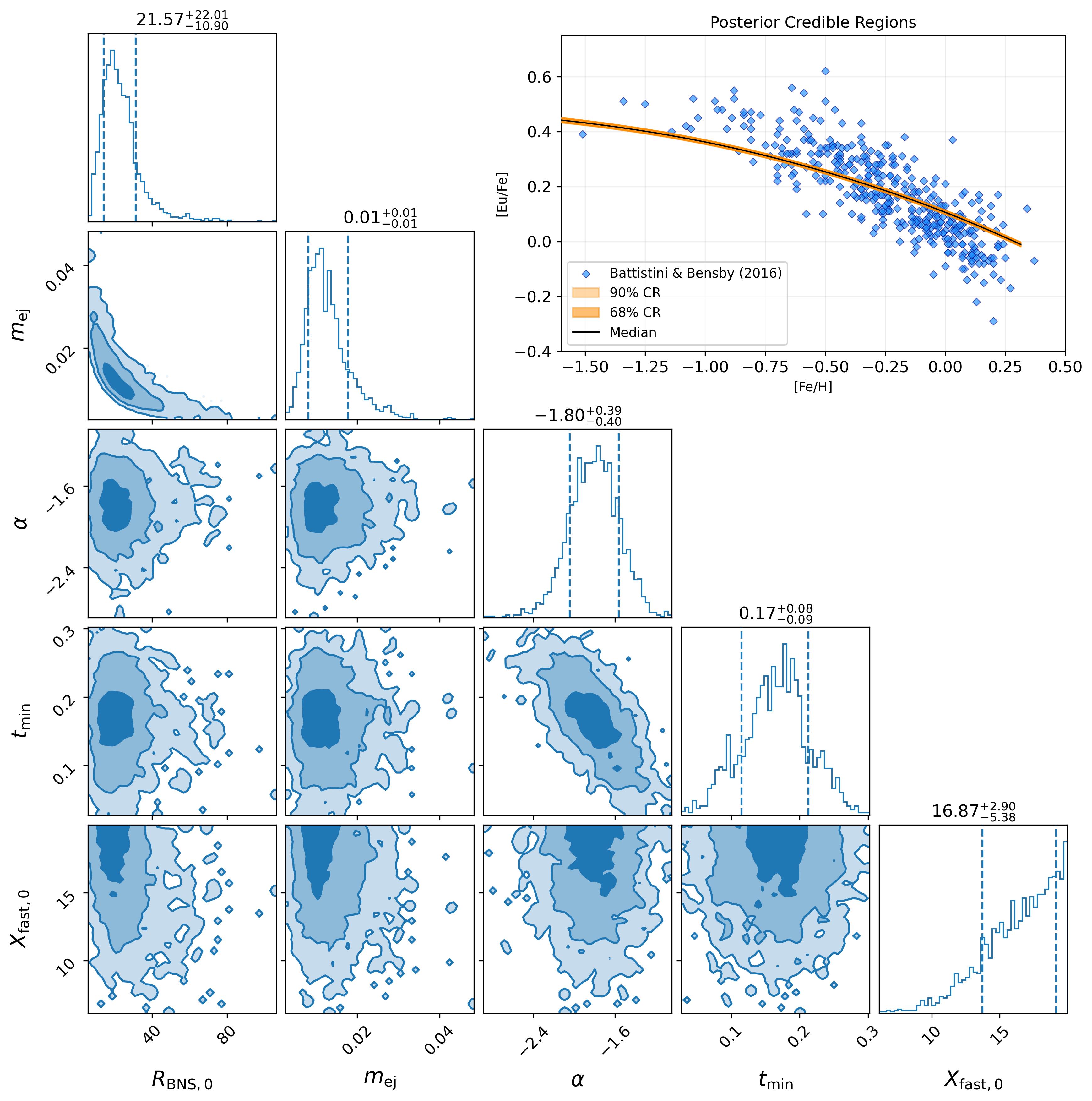}
    \caption{\textbf{Current GW searches miss fast-merging BNSs.} Diagonal panels show marginalized posteriors (medians and 68\% credible intervals); off-diagonal panels show joint posteriors (filled credible regions). Parameters are the delayed-channel BNS merger rate $R_{\rm BNS}$ (${\rm Gpc}^{-3}\, {\rm yr}^{-1}$) (here referring to the \emph{delayed} component only), the common ejecta mass $m_{\rm ej}$ (${\rm M}_\odot$), the delayed-channel delay-time parameters $\alpha$ and $t_{\rm min}$ (${\rm Gyr}$), and the scaling factor $X_{\rm fast}$ that sets the fast-merging rate as $R_{\rm fast} = X_{\rm fast}\,R_{\rm BNS}$. The subscript `0' in the parameter labels indicate that the quantities are evaluated in the local universe ($z=0$). The large panel at upper right displays the posterior-predicted credible regions (orange) in $[\mathrm{Eu}/\mathrm{Fe}]$ versus $[\mathrm{Fe}/\mathrm{H}]$, with Milky Way disk measurements from \cite{Battistini} shown as blue points.}
    \label{fig:fig2_corner_plus_track}
\end{figure*}

In this appendix, we present additional plots that shed light on further details of each of the three analyses discussed in the main text. Figure~\ref{fig:fig1_corner_plus_track} shows corner plots of the posterior distributions for all model parameters: the total BNS rate $R_{\rm BNS}$ (including both delayed and fast-merging channels); the ejecta mass $m_{\rm ej}$ (i.e., the per-event $r$-process yield, assumed common to both channels); the delay-time parameters $\alpha$ and $t_{\rm min}$ of the delayed BNS channel; and the fraction $f_{\rm fast}$, which specifies the portion of $R_{\rm BNS}$ assigned to the fast-merging channel required to reproduce the observed abundance trends of Milky Way disk stars~\citep{Battistini}. The large panel in the upper right shows the 90\% credible regions (orange) in the $[\mathrm{Eu}/\mathrm{Fe}]$--$[\mathrm{Fe}/\mathrm{H}]$ plane implied by the posterior samples; observed abundances from \cite{Battistini} are overplotted as blue points.
Unless otherwise stated, the axes in the corner plots span the full prior ranges used in the corresponding inference. In the hidden-fast-merger analysis, $X_{\rm fast}$ is assigned a finite prior range chosen to cover the physically relevant rate enhancements while keeping the numerical sampling well behaved. The posterior in Fig.~\ref{fig:fig2_corner_plus_track} lies close to the high-$X_{\rm fast}$ side of this range, indicating that the abundance data would tolerate even larger hidden prompt contributions if they were permitted. This behavior is one reason we interpret the model as requiring an extreme and astrophysically disfavored hidden population rather than as providing evidence for a specific value of $X_{\rm fast}$.

Fig.~\ref{fig:fig2_corner_plus_track} presents the similar details for the analysis discussed in Fig.~\ref{fig:ecc_case2_summary}, in which $R_{\rm BNS}$ refers exclusively to the \emph{delayed} BNS channel. The ejecta mass $m_{\rm ej}$ is again taken to be common, and the fast component is encoded via a multiplicative factor $X_{\rm fast}$ such that $R_{\rm fast}=X_{\rm fast}\,R_{\rm BNS}$.

\begin{figure*}
    \centering
    \includegraphics[width=0.75\textwidth]{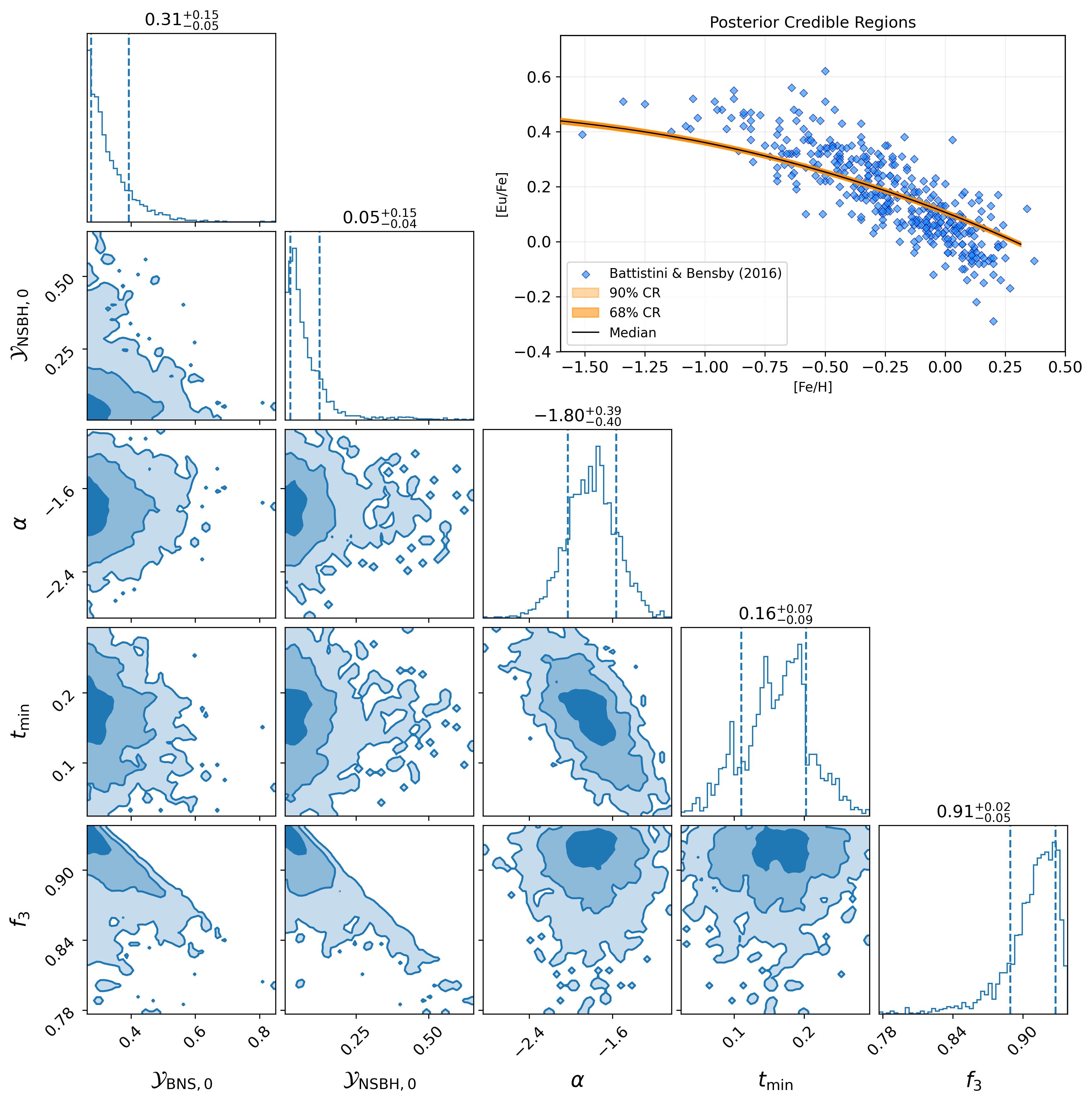}
    \caption{\textbf{NSBH Mergers as a Secondary Channel.} Diagonal panels show marginalized posteriors (medians and 68\% credible intervals); off-diagonal panels show joint posteriors (filled credible regions). Parameters are the rate times mass ejecta products $\mathcal{Y}_{\rm BNS}=R_{\rm BNS} m_{\rm ej}^{\rm BNS}$ and $\mathcal{Y}_{\rm NSBH}=R_{\rm NSBH} m_{\rm ej}^{\rm NSBH}$ (${\rm M}_\odot\, {\rm Gpc}^{-3}\, {\rm yr}^{-1}$), the BNS delay-time parameters $(\alpha, t_{\min})$, and the fractional third-channel contribution $f_{3}$ defined via $\mathcal{Y}_{3}=f_{3}\,\mathcal{Y}_{\rm tot}$. The subscript `0' in the parameter labels indicate that all these quantities are evaluated in the local universe ($z=0$). The large panel at upper right displays the posterior-predicted credible regions (orange) in $[\mathrm{Eu}/\mathrm{Fe}]$ versus $[\mathrm{Fe}/\mathrm{H}]$, with Milky Way disk measurements from \cite{Battistini} shown as blue points.}
    \label{fig:nsbh_corner_plus_track}
\end{figure*}

Finally, Fig.~\ref{fig:nsbh_corner_plus_track} summarizes the analysis in which we include both BNS and NSBH contributions with an agnostic third channel to capture any residual enrichment not accounted for by them. The sampled parameters are the rate times mass ejecta products $\mathcal{Y}_{\rm BNS}\!\equiv\!R_{\rm BNS}\,m_{\rm ej}^{\rm BNS}$ and $\mathcal{Y}_{\rm NSBH}\!\equiv\!R_{\rm NSBH}\,m_{\rm ej}^{\rm NSBH}$, the BNS channel delay-time parameters $(\alpha, t_{\min})$, and a fractional contribution parameter $f_{3}$ that encodes the third channel's share of the total mass-injection rate (we take $\mathcal{Y}_{3}=f_{3}\,\mathcal{Y}_{\rm tot}$ with $\mathcal{Y}_{\rm tot}=\mathcal{Y}_{\rm BNS}+\mathcal{Y}_{\rm NSBH}+\mathcal{Y}_{3}$). The large panel at upper right again shows the posterior-predicted credible regions in the $[\mathrm{Eu}/\mathrm{Fe}]$--$[\mathrm{Fe}/\mathrm{H}]$ plane with the same observational sample overplotted.

\section{Impact of star formation history models}\label{app:sfr}

\begin{figure*}
    \centering
    \includegraphics[width=0.32\textwidth]{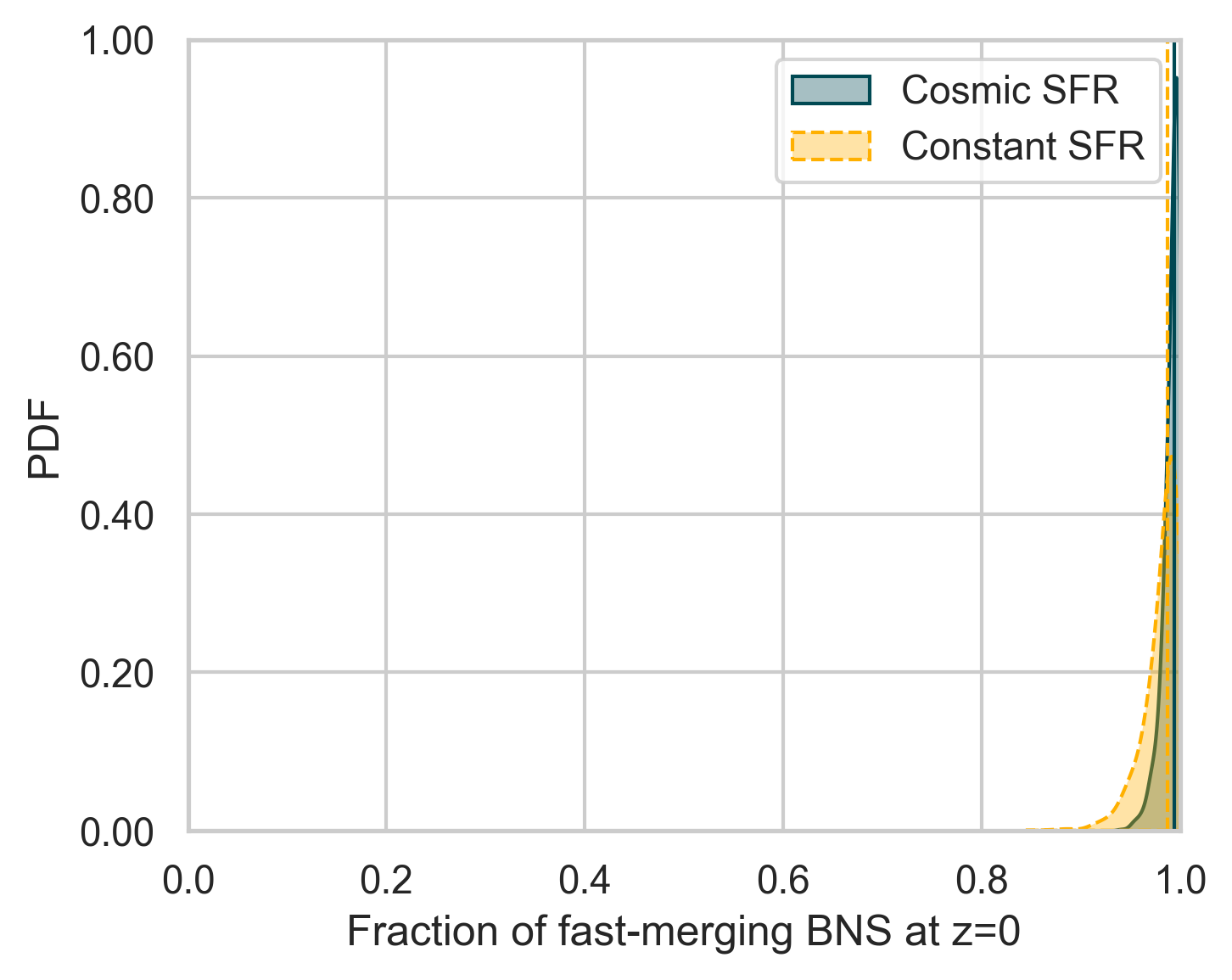}
    \includegraphics[width=0.32\textwidth]{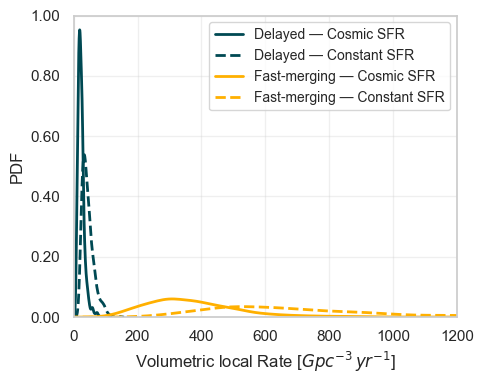}
    \includegraphics[width=0.32\textwidth]{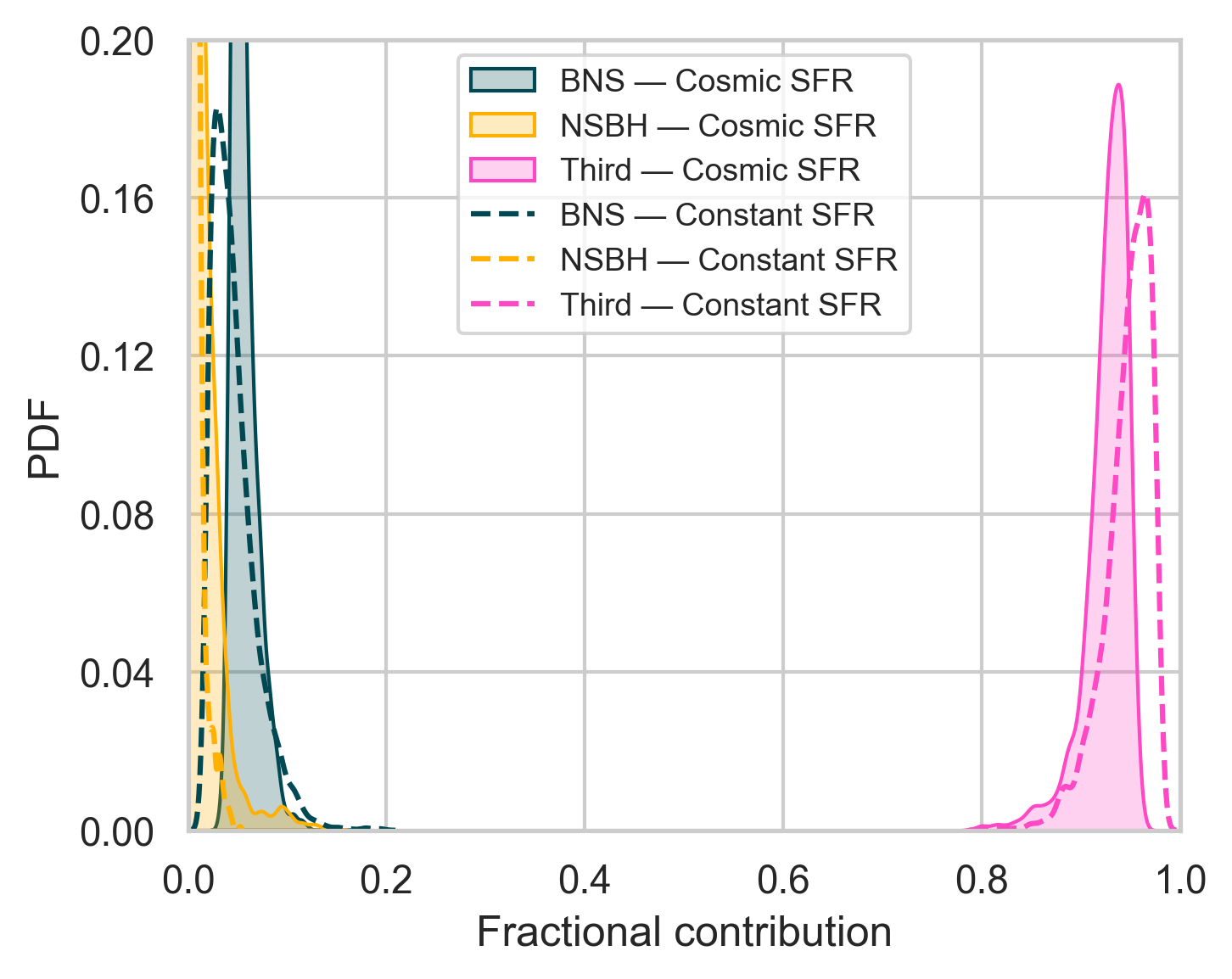}
    \caption{Sensitivity to the star-formation history (SFR model). We compare the main analysis that adopts the cosmic SFR (solid curves) with a constant-SFR as an extreme case (dashed). [Left] Fast-merging BNS as a secondary channel (in addition to delayed BNS), assuming they are detectable by current GW searches. The figure shows the fraction of fast-merging BNS in the GW-inferred local BNS rate needed to reproduce the observed r-process abundance. [Middle] Fast-merging BNS are assumed to be missed by current GW searches. The figure shows the required local volumetric rates for the delayed and fast-merging BNS channels. [Right] NSBH mergers as the second channel. The figure is the same as Fig.~\ref{fig:nsbh_case1_summary} with two SFR models.}
    \label{fig:sfr-effect}
\end{figure*}

In this section, we assess the sensitivity of our results to the assumed SFR model. We repeat the full inference under a \emph{constant SFR}---an extreme limiting case---and compare it to the \emph{cosmic SFR} used in the main text. The three panels in Fig.~\ref{fig:sfr-effect} summarize the outcome for each secondary-channel scenario considered in the paper: (i) fast-merging BNS assumed detectable by current GW searches; (ii) fast-merging BNS assumed missed by current searches; and (iii) NSBH mergers. Across all three cases the posterior shifts between the two SFR assumptions are modest, and the qualitative conclusions of the paper are unchanged: neither fast-merging BNS nor NSBH mergers can account for the Milky Way's $r$-process enrichment. Bayesian model selection decisively prefers the cosmic SFR in every scenario, with logarithmic Bayes factors $29.98$, $29.34$, and $29.67$, respectively, for the three cases. This justifies adopting the cosmic SFR in the main analysis and demonstrates that our main findings are robust to the SFR choice.

\bibliography{main}{}
\bibliographystyle{aasjournalv7}

\end{document}